# Optimization and analysis of large scale data sorting algorithm based on Hadoop

Zhuo Wang, Longlong Tian, Dianjie Guo, Xiaoming Jiang
Institute of Information Engineering, Chinese Academy of Sciences
{wangzhuo, tianlonglong, guodianjie, jiangxiaoming}@iie.ac.cn

## Abstract

When dealing with massive data sorting, we usually use Hadoop which is a framework that allows for the distributed processing of large data sets across clusters of computers using simple programming models. A common approach in implement of big data sorting is to use shuffle and sort phase in MapReduce based on Hadoop. However, if we use it directly, the efficiency could be very low and the load imbalance can be a big problem.

In this paper we carry out an experimental study of an optimization and analysis of large scale data sorting algorithm based on hadoop. In order to reach optimization, we use more than 2 rounds MapReduce. In the first round, we use a MapReduce to take sample randomly. Then we use another MapReduce to order the data uniformly, according to the results of the first round. If the data is also too big, it will turn back to the first round and keep on. The experiments show that, it is better to use the optimized algorithm than shuffle of MapReduce to sort large scale data.

## 1 Introduction

### 1.1 Hadoop

The Apache Hadoop software library is a framework that allows for the distributed processing of large data sets across clusters of computers using simple programming model [1]. It is designed to speed up from single servers to hundreds of or thousands of machines, each offering local storage and computation. Rather than rely on hardware to deliver high-availability, the library itself is designed to detect and handle failures at the application layer, so delivering a highly-available service on top of a cluster of computers, each of which may be prone to failures.

### 1.2 MapReduce

MapReduce is a programming model and an associated implementation for processing and generating large data sets [2]. A *map* function can process a key/value pair to generate a set of intermediate key/value pairs, and a *reduce* function merges all intermediate values associated with the same intermediate key. This model can express many tasks in our real world.

Programs written in this kind of functional style can be automatically parallelized and executed on a large cluster of commodity machines. The run-time system deals with the details of scheduling the program's execution across a set of machines and managing the required inter-machine communication, partitioning the input data, handling machine failures. This let programmers without any experience with distributed systems and parallel to easily use the resources of a large distributed system.

MapReduce makes sure that the input to every reducer is sorted by key. The process by which the system executes the sort and transfers outputs of the map to the reducers as inputs is known as shuffle. In many ways, the shuffle is the heart of MapReduce.

Shuffle is the process of transferring data from the mappers to the reducers. Sorting in shuffle saves time for the reducer, helping it easily differentiate when a new reduce task should start. It begins a new reduce task simply, when the following key is different from the previous one, to put it easily.

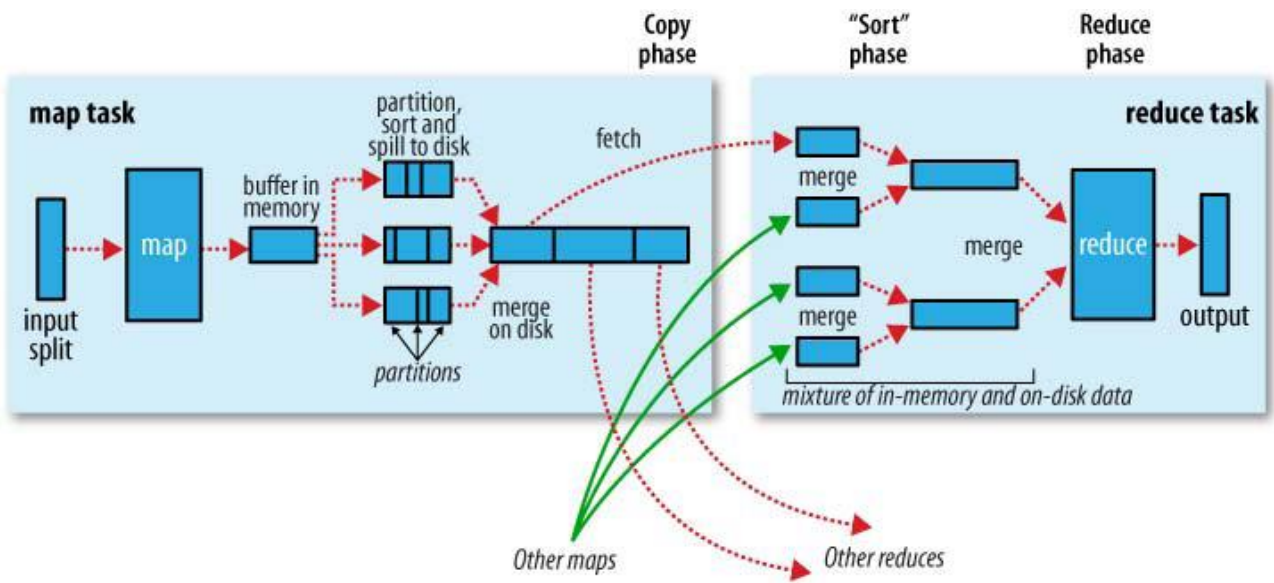

**Figure 1-1** the process of MapReduce

### 1.3 Optimization of the data sorting algorithm

As the efficiency of shuffle may be low, we can achieve optimization based on Hadoop by some operating. In our work, we decide to use more than 2 rounds MapReduce. Because some nodes in cluster may take very heavy jobs while the others have basically no work to do. So we first to divide the data to be sorted. In the first round, we use a MapReduce to take sample randomly. After it, we almost know the distribution of the data. We can create some new files, every of which has average data .Then we use another MapReduce to order the data uniformly, according to the results of the first round. If the data is also too big, it will turn back to the first round to be divided and keep on. The experiments show that, it is better to use the optimized algorithm than shuffle of MapReduce to sort large scale data.

## 2 Design and implementation

### 2.1 Basic idea of the algorithm

The basic idea is that the data is divided into inter-block ordered data, so that it can be sorted in the memory as much as possible. Figure 2-1 shows the general design of this algorithm.

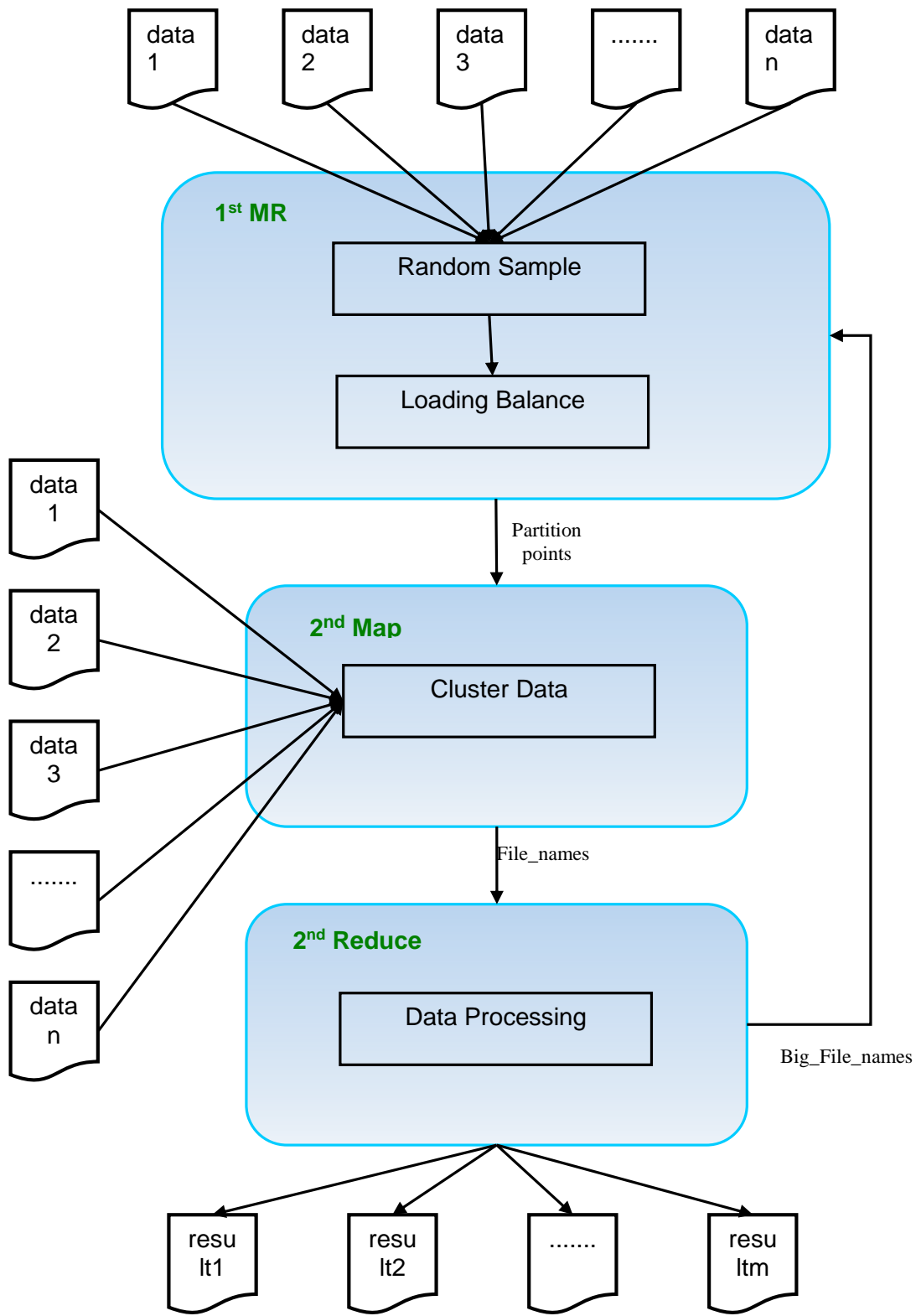

**Figure2-1** the process of this algorithm

I division and sort based on files

(1)Get the distribution of data to be sorted, and set the division site.

First obtain the distribution information of the files. Take three sites of data to make sample for each file to be sorted, and sample 4KB data for each site, using the style like <data, the number of data > to save it in HashMap. The data in map will be sorted by priority queue after statistics.

Set division site, where the total number of sampling data is counted that is named sampleCount; obtained the total length of all files to be sorted by the meta information of files and the length is named totalLength; set the size of data to be sorted on every reduce manually, and the size is named blockSize; and set the length of divided segment is divideNums, so we can know that totalLength divided by sampleCount equals blockSize divided by divideNums,divideNums equals sampleCount multiplied by blockSize divided by totalLength, combining with priority queuing we can be obtained classified site. Set the number of reduce by adding the number of classified site and 1, and we can set the threshold value as a reference, whichever is the lowest value will be the number of reduce.

(2) For each process of the *Map*, we can obtain the information of classified site through the process of setup; initialize the middle file, and generate n (the number of classified site + 1) files; Generate files using a random number, if there is a conflict, then re-generate the file name. When processing each data, data is stored in different files according to classified site; Wait for the end of writing files after the data processing; take the dividing data segment of each file as a key, the file path and file name as a value; Send to the reduce according to the partition function.

(3) Use the number of key module reduce to send out map into reduce in partition process.

(4) After obtaining <key, value> in reduce, we know what the corresponding divided data segment is and the corresponding data segments are; Obtain the meta information of these files, the total length of the statistics, if it is greater than the threshold value in RAM, then write the serial number of corresponding data in output, and return with doing nothing; Otherwise, the contents of each data file is read and placed in the priority queue for sorting in memory, write the result into a file named the number of data segment, and put in the result directory, such as */result/0*.That is to say, if the output directory is */result/*, then corresponding number of data segment is 0.

II divide and sort again based on files

(1)Read the contents of the output files and save all of the unsorted data segment numbers.

(2)Take the data corresponding the middle file as a target for each unsorted segment number; Repeat the procedure1 again; Generate file like */result/1_2*, and 1 is the first unsorted data bit segment, 2 is the data bit segment corresponding sort in process 2.

III division and sort based on datasets.

For the second unsorted bit segment, send the data in different interval to different reduce in accordance with different partition method after generating the division by classifying site in process 1; Write the results into directory after automatically sorting in shuffle, which corresponds to the format of the form */result/1_2_3*, 1 and 2 are the first two unsorted data corresponding bit segment, 3 is the sorted bit segment data.

## 2.2 concrete implementation of the algorithm

The concrete processes executes as follows.

(1) Take sample simply and divide the datasets. We can suppose that the size of whole datasets is 100M and the threshold of the division is 20M. We can get that the count of the sample is 100 and the number of divisions equals 20. So the situation of division is showed in **Figure 2-2**. And the number of reducer is 6.

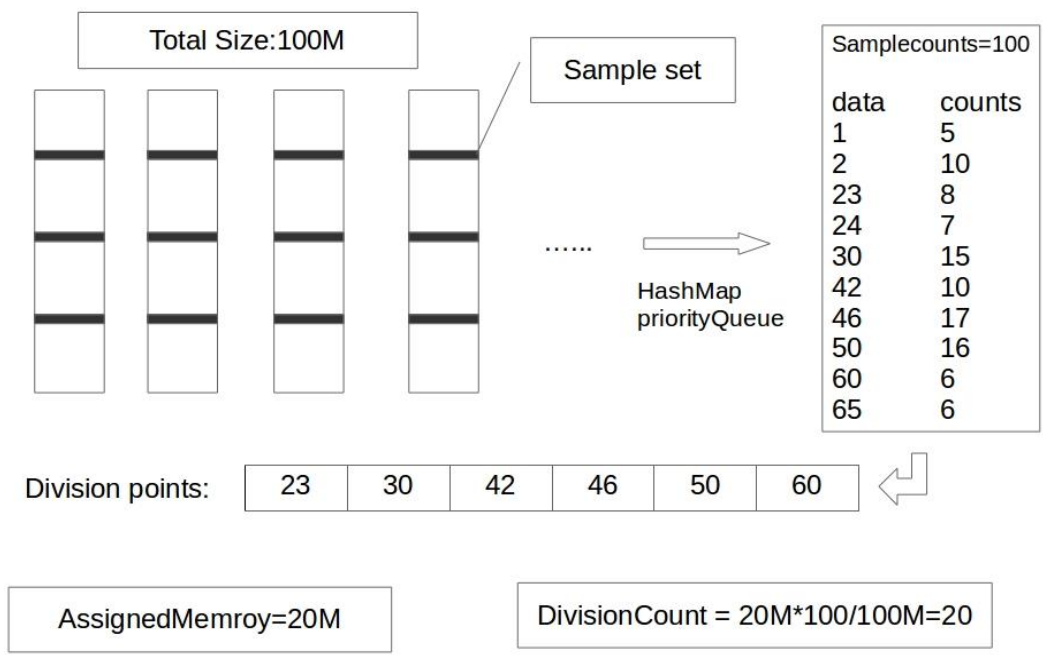

**Figure 2-2** take sample and divide the dataset

(2) Implement the *setup* function in the *Mapper* and the intermediate results are saved in the directory *middle*. The different filenames corresponding to its scope of data. For example, the data number 23 corresponding to the file *Middle/0/_0.23568* and the data is larger than 23 and smaller than 42 corresponding to the file *Middle/1/_0.54668* and so on. This step is showed in **Figure 2-3**.

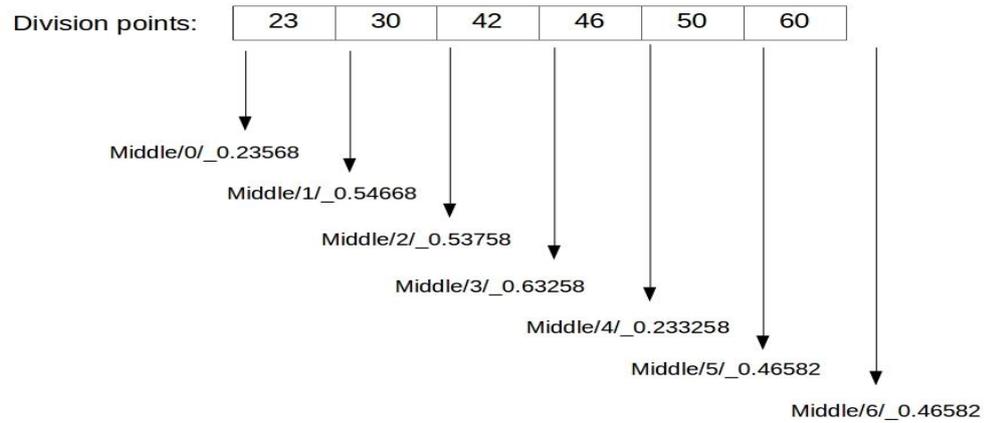

**Figure 2-3** the filenames corresponding the scope of data

(3) **Figure 2-4** shows the process that *map* generate the different division file. If *map2* and *map1* generate the same *filename*, we need to generate filenames again till there is not conflict in filenames.

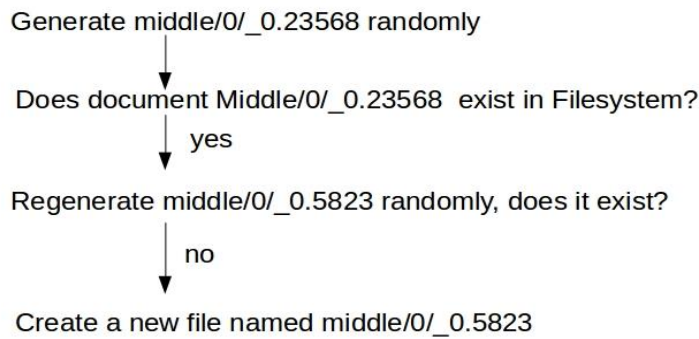

**Figure 2-4** the process that map generate filenames

(4) **Figure 2-5** shows the content of different file generated in step 3.

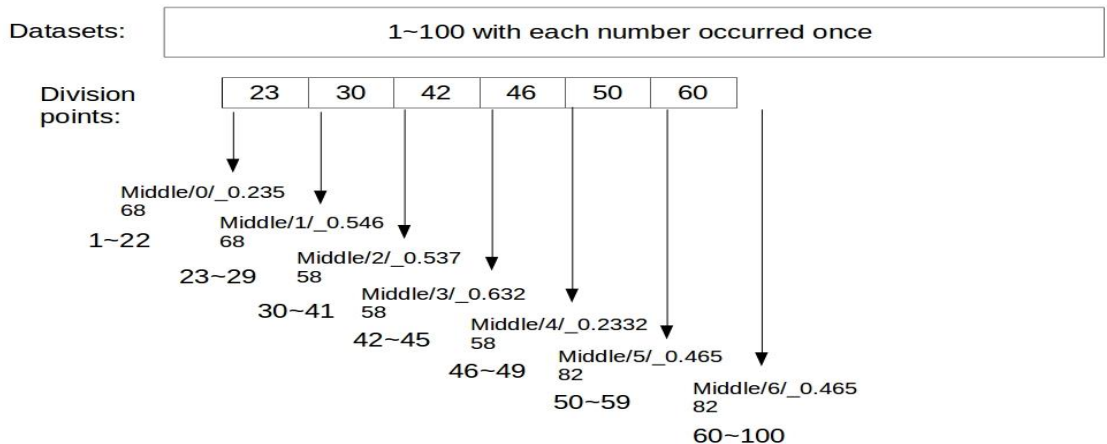

**Figure 2-5** the content of the intermediate files

(5) Send the number of data scope and the corresponding directory to the *reducer*. We implement this by a function that the number modules the number of *reducer*.

(6) **Figure 2-6** and **Figure 2-7** show that the size of file is smaller than 20M and write the files into the directory. Otherwise, return the data directly.

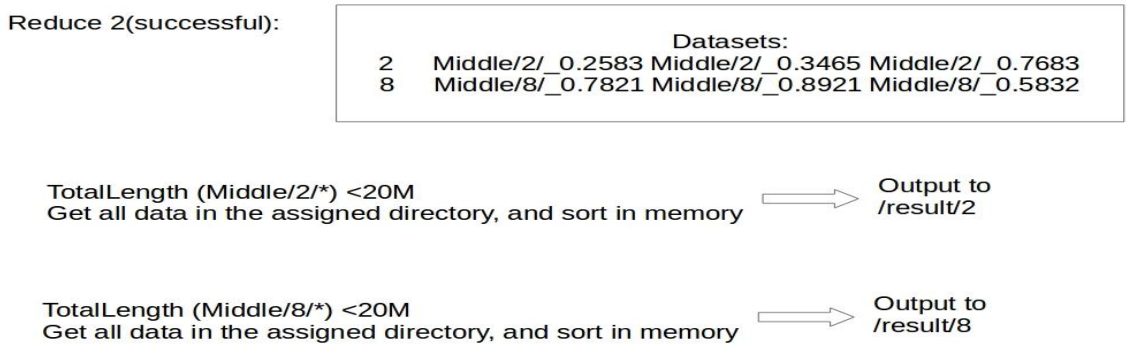

**Figure 2-6** size of file is smaller than 20M

```
Reduce 2(failed):                Datasets:
                          2  Middle/2/_0.2583 Middle/2/_0.3465 Middle/2/_0.7683
                          8  Middle/8/_0.7821 Middle/8/_0.8921 Middle/8/_0.5832

TotalLength (Middle/2/*) >20M    ⇒  Output to
                                    /part-r-00002
                                    2

TotalLength (Middle/8/*) >20M    ⇒  Output to
                                    /part-r-00002
                                    8

/output/part-r-00002 data:
2
8
```

**Figure 2-7** size of file is larger than 20M

(7) The process ends when there isn't number of data scope outputting. Otherwise, repeat from step 1 to step 6.

(8) The final result can be seen in **Figure 2-8**. The number of MapReduce process depends on the precision which the sample represent the whole datasets.

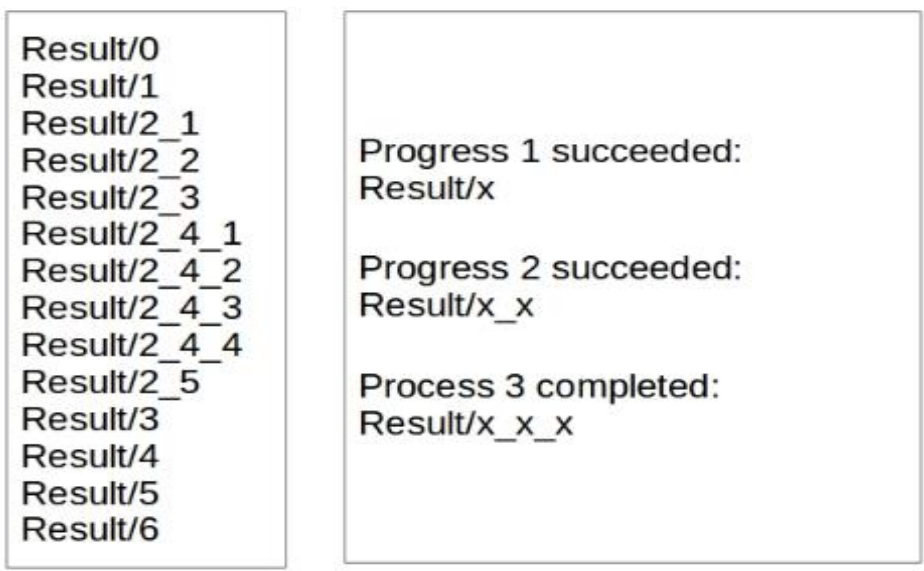

**Figure 2-8** the final result

# 3 Results and analysis

We have implemented this algorithm with Java. We have already run this algorithm on the pseudo-distributed mode where each Hadoop daemon runs in a separate Java process. The program sorts several datasets with different size and distribution. Then we sort the same datasets with the process of shuffle. The **Table 3-1** shows the result of comparison.

| size | baseline(s) | new_partition(s) |
|---|---|---|
| **30M** | 41 | 42 |
| **40M** | 45 | 45 |
| **50M** | 62 | 50 |
| **60M** | 58 | 52 |
| **70M** | 57 | 58 |
| **80M** | 62 | 56 |
| **90M** | 74 | 60 |
| **100M** | 77 | 65 |
| **110M** | 81 | 76 |
| **120M** | 88 | 80 |
| **130M** | 96 | 90 |
| **140M** | 97 | 88 |
| **150M** | 112 | 100 |
| **180M** | --- | 140 |

**Table 3-1** the running time of two processes

**Table 3-1** shows that out algorithm performs better than sort process of shuffle. We generate the testing data randomly. And these situations can represent the distribution of normal datasets. We can also find that the program executes the process 2 barely. It means that the sample is typical for the whole datasets. The sort process of shuffle cannot work well when the size of input data is larger than 180M. The pseudo-distributed mode we used doesn't possess enough memory to process surplus data. However, the algorithm we design can work well because this process avoid to utilize the shuffle and try to sort data in the main memory. We cannot test larger datasets due to the memory limit of the memory. But the algorithm we design indeed performs better than the sort process of shuffle.

# 4 Conclusion and future work

The contribution of this paper is a new strategy which is suitable for large-scale data sorting and load-balancing based on Hadoop. Considering the load-imbalance leading by the assigning jobs of MapReduce, we take samples simply of the datasets and get the information about its distribution basically. We partition the original datasets averagely to make the work of each Reduce jobs approximately equivalent.

In the process of MapReduce, there is a process named shuffle between Mapper and Reducer and shuffle can sort the output of Mapper by the key. The key-value pairs are written into the local file system—disks, and sent to the same Reducer if the output has the same key. Therefore, shuffle can make the datasets sorted. By comparing the efficiency that the shuffle process sort the datasets, we can find that this algorithm performs better than the sort processing of shuffle in some occasions. We will perform this algorithm on full-distributed model consist of clusters and test large scale datasets to verify the correctness of this algorithm.

We are about to improve this algorithm continuously. We have already learned other paralleling models such as Spark, Pregel, and GraphLite. Spark is an architecture for fast and general data processing on large clusters [3]. Pregel and GraphLite are systems built for large scale graph processing [4]. These models perform well on the situations that exists many iterative steps. We will modify this algorithm to adapt to these models.